# A cascaded dual-domain deep learning reconstruction method for sparsely spaced multidetector helical CT

Ao Zheng[1,2], Hewei Gao[1,2], Li Zhang[1,2] and Yuxiang Xing[1,2(✉)]

[1] Department of Engineering Physics, Tsinghua University, Beijing, 100084, China
xingyx@mail.tsinghua.edu.cn

[2] Key Laboratory of Particle & Radiation Imaging (Tsinghua University), Ministry of Education, China

**Abstract.** Helical CT has been widely used in clinical diagnosis. Sparsely spaced multidetector in z direction can increase the coverage of the detector provided limited detector rows. It can speed up volumetric CT scan, lower the radiation dose and reduce motion artifacts. However, it leads to insufficient data for reconstruction. That means reconstructions from general analytical methods will have severe artifacts. Iterative reconstruction methods might be able to deal with this situation but with the cost of huge computational load. In this work, we propose a cascaded dual-domain deep learning method that completes both data transformation in projection domain and error reduction in image domain. First, a convolutional neural network (CNN) in projection domain is constructed to estimate missing helical projection data and converting helical projection data to 2D fan-beam projection data. This step is to suppress helical artifacts and reduce the following computational cost. Then, an analytical linear operator is followed to transfer the data from projection domain to image domain. Finally, an image domain CNN is added to improve image quality further. These three steps work as an entirety and can be trained end to end. The overall network is trained using a simulated lung CT dataset with Poisson noise from 25 patients. We evaluate the trained network on another three patients and obtain very encouraging results with both visual examination and quantitative comparison. The resulting RRMSE is 6.56% and the SSIM is 99.60%. In addition, we test the trained network on the lung CT dataset with different noise level and a new dental CT dataset to demonstrate the generalization and robustness of our method.

## 1 Introduction

Over the past decades, helical CT has been widely used in clinical diagnosis. Compared to MRI, it has a lot of advantages, such as, short scanning time and high resolution in bones. However, the radiation dose are our main concerns. Since helical CT has been invented, many reconstruction methods are promoted to successively improve the quality of reconstructed images.

Earlier algorithm for helical CT were filtered back-projection (FBP) [1], but it was approximate. In 2002, Katsevich [2] first proposed a theoretically exact FBP formula for helical CT. It was initially developed for helical cone-beam geometry and was later extended to general trajectories [3,4]. Inspired by [2], an alternative approach for theoretically exact reconstruction method was also proposed in the back-projection filtration (BPF) format [5]. For lowing the radiation dose, people intend to raise pitch and shorten scanning time. But these algorithms seem not work well. Aiming at high pitch in helical CT, PI-Methods [6] are a series of algorithm which only consider the projection data within the Tam window and rebinning them to oblique parallel beam. It shows a convincing result in high pitch situation.

When projection data is noisy, instead of analytical reconstruction methods, iterative reconstruction methods, such as, SIR [7], MBIR [8] are adopted. In addition, we could also add punishment based on priori hypotheses. According to Compress Sensing [9], [10], we can add total variation (TV) of the image as a constraint. Different from TV, q-Generalized Gaussian Markov random field (qGGMRF) [11] supposes the distribution of attenuation is local smooth. But this hypothesis may induce low quality of details. For solving this problem, non-local means filter (NLM) [12] extends the neighborhood to a large search window and uses a more complicated weighting method. Although the results of iterative reconstruction methods are often of high quality, it will spend a huge computational cost which is unacceptable in practice.

Recently, in medical imaging, CNN has shown its great potential on classification, detection and segmentation. And in CT area, there are many methods based on CNN focusing on reconstruction. Some of them use CNN to directly reconstruct from sinogram. For example, [13] proposed a deep learning reconstruction method which contains data acquisition conditions in network parameters. Others utilize analytical reconstruction-based operator to change data from projection domain to image domain and use deep learning methods to solve ill-condition problems. [14] used U-Net [15] in image domain to solve sparse-view reconstruction problem in parallel beam CT. The U-Net combines multiresolution features and learns in a residual way, which achieved promising results. From image domain to projection domain, [16] utilized cascaded CNNs to reconstruct in sparse-view fan-beam situation. And it demonstrated that two CNN in both projection domain and image domain perform better than image domain CNN only. For 3D problem, [17] used U-Net in dual-domain to solve the low-dose problem in cone-beam CT. It proposed a slice-wise reconstruction method which outperformed analytical reconstruction methods. These works imply that CNN has a promising potential to solve CT reconstruction problems with insufficient data.

In this work, we employ sparsely spaced multidetector in z direction which can speed up scan and reduce the cost of detectors. It can also lower radiation dose by appropriately setting collimators. However, based on [18], in real system, there are more factors to be considered. Therefore, in this paper, we are not setting collimators. In our system, the projection data is insufficient for reconstruction in general concern. For this kind of problem, analytical reconstruction methods naturally result in severe artifacts and model-based iterative reconstruction could be used to cope with the ill-condition of this problem, but a huge computational cost is unavoidable. Based on the former research [17] of our group, we propose a reconstruction method using cascaded CNN learning both data transform in projection domain and error reduction in image domain. Our method not only shows an encouraging result in test datasets but also performs robustly in generalization.

## 2 Method

In this section, we introduce the geometry of our system and data preprocessing first. Then we explain our deep learning reconstruction network. At last, we will demonstrate the detailed network architecture and the whole training procedures.

### 2.1 System geometry and data preprocessing

We denote $\mathbf{p}^{\text{Heli}} \in \Re^{C\times R\times A}$ be the complete helical projection data with a detector of $C$ columns, $R$ rows and $A$ scanning angles in total. As shown in Fig. 1(a), the sparsely spaced helical CT we are interested only equipped with a small number of detector rows $R' \ll R$, and we are to reconstruct a volume from sparse helical CT data $\mathbf{p}^{\text{sp}} \in \Re^{C\times R'\times A}$. Without losing any generality, we use our method to reconstruct one slice for illustration. In practice, we can reconstruct slice by slice to form a whole 3D volume. Let $\boldsymbol{\mu}(z)$ be the slice to be reconstructed, and the view angle corresponding to this axial location is $\theta(z)=\beta z$. It is straightforward to locate data relevant to this slice, denoted by $\left\{\mathbf{p}^{\text{Heli}}\right\}_{\in\boldsymbol{\mu}(z)}$ in non-sparse case and $\left\{\mathbf{p}^{\text{sp}}\right\}_{\in\boldsymbol{\mu}(z)}$ in our sparse−detector scan system. Fig. 1(b) shows the illustration in non-sparse system. We assume the view angle coverage of $\left\{\mathbf{p}^{\text{Heli}}\right\}_{\in\boldsymbol{\mu}(z)}$ by $\left[\theta_1, \theta_2\right]$ with $\theta_1 < \theta(z) < \theta_2$. As shown in Fig. 1(c), for $\boldsymbol{\mu}(z)$ in arbitrary view angles, we have $\Delta z = (\theta - \theta(\text{z}))/\beta,\quad \forall\theta \in \left[\theta_1, \theta_2\right]$. Suppose the thickness of detector rows and the radius of reconstructed field of view are $\Delta s$ and $h$ respectively. Hence, data from detector rows within $[R_1(\theta),\ R_2(\theta)]$:

$$\begin{aligned} R_1(\theta) &= \frac{R}{2} + \frac{|\mathbf{SD}|}{|\mathbf{SO}| - (-1)^{\gamma}\cdot h}\frac{\theta-\theta(z)}{\beta\Delta s} \\ R_2(\theta) &= \frac{R}{2} + \frac{|\mathbf{SD}|}{|\mathbf{SO}| + (-1)^{\gamma}\cdot h}\frac{\theta-\theta(z)}{\beta\Delta s} \end{aligned},\ \gamma = \begin{cases} 0 & \theta < \theta(z) \\ 1 & \theta > \theta(z) \end{cases},\ \theta \in [\theta_1,\ \theta_2] \qquad (1)$$

are the data shall be used to reconstruct the slice of interest, which is shown as the light gray area in Fig. 1(d). Because of the specificity of helical CT, $R_2(\theta) - R_1(\theta)$, the number of detector rows detecting $\boldsymbol{\mu}(z)$ varies in different view angles. For convenience, we choose the maximum of $R_2(\theta) - R_1(\theta)$ to form a cubic data truck, $\left\{\mathbf{p}^{\text{Heli}}\right\}_{\in\boldsymbol{\mu}(z)} \in \Re^{C\times\theta\times r},\ \ \theta \in [\theta_1, \theta_2]$ and $r = \lceil \max(R_2(\theta) - R_1(\theta)) \rceil$. Here, $\lceil\ \ \rceil$ means round toward positive infinity. In practice, we firstly use linear interpolation to convert $\left\{\mathbf{p}^{\text{sp}}\right\}_{\in\boldsymbol{\mu}(z)}$ to $\left\{\hat{\mathbf{p}}^{\text{Heli}}\right\}_{\in\boldsymbol{\mu}(z)}$ and then correct the interpolation error in following steps. Therefore, the input of our network in sparse−detector scan system is $[\hat{\mathbf{p}}^{\text{Heli}}_{\text{sp}}(z)]_{C,\theta,r}$.

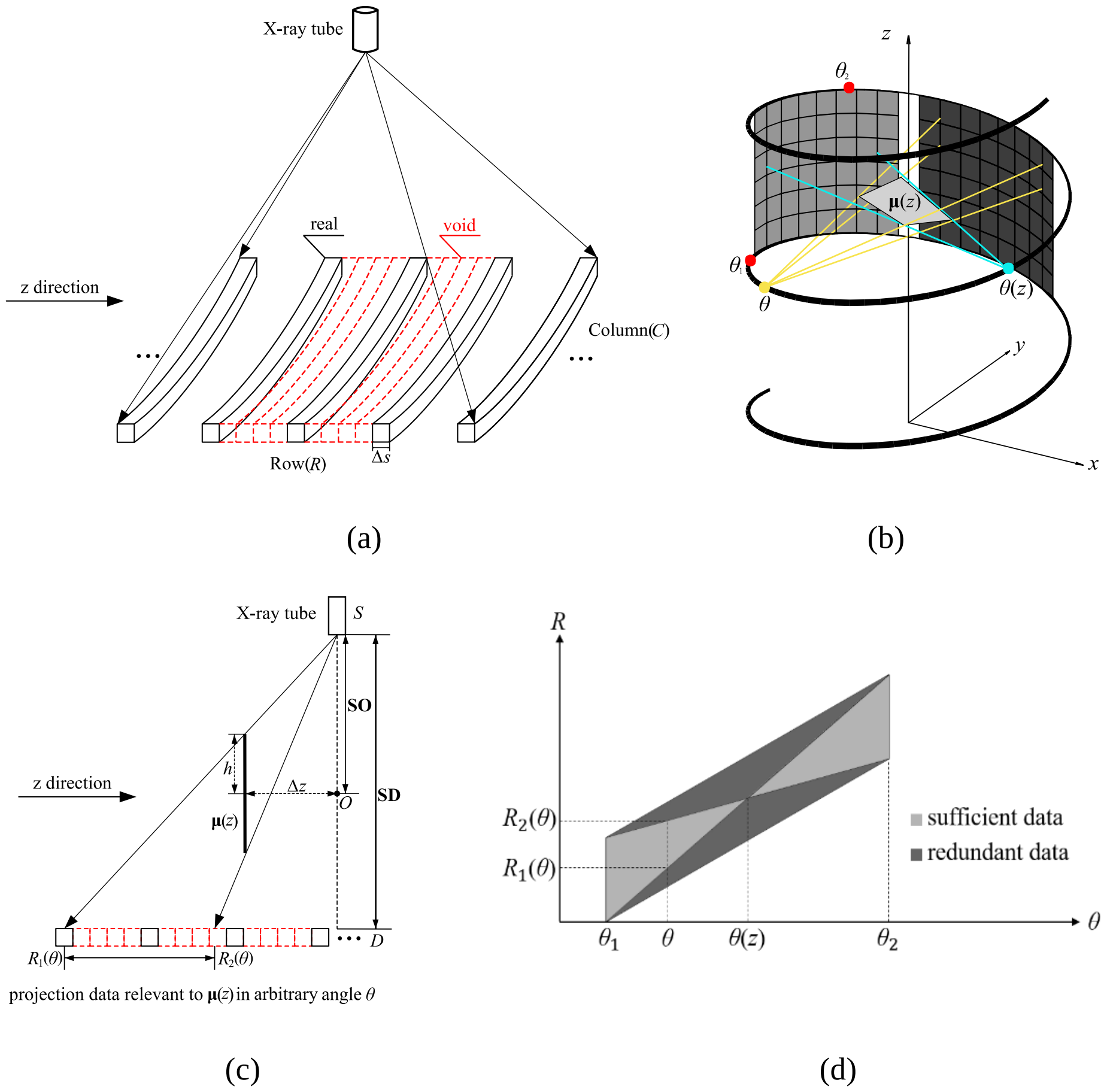

**Fig. 1.** illustrations of our system and helical projection data.

## 2.2 Deep learning reconstruction network.

The overall reconstruction network is composed of three parts which can be described as follows.

**Projection Domain CNN.** The CNN in projection domain aims at recovering helical projection and converting it to slice-independent fan-beam projection data, which is to suppress helical artifacts and reduce the following computational cost.

The input of projection domain CNN is $\hat{\mathbf{p}}_{\text{sp}}^{\text{Heli}}(z)$. The targeted output is virtual fan-beam projection data denoted as $\hat{\mathbf{p}}$. It is convenient to set the fan-beam projection to be same as $\hat{\mathbf{p}}_{\text{sp}}^{\text{Heli}}(z)$ except of one row. The ground truth label $\mathbf{p}^*$ can be obtained by projecting high-quality label images $\boldsymbol{\mu}^*$:

$$\mathbf{p}^* = \mathbf{H}^{\text{fan}} \boldsymbol{\mu}^* \tag{2}$$

with $\mathbf{H}^{\text{fan}}$ being the system matrix of fan-beam projection.

**Analytical reconstruction layer.** Since $\hat{\mathbf{p}}$ is a fan-beam projection, we can construct a layer as a straightforward analytical reconstruction operator such as FBP to transfer data from projection domain to image domain, i.e. inverse Radon. It contains three sub-steps, weighting, filtering and back-projection. Denote the output of this inverse Radon layer as $\boldsymbol{\mu}_{\text{FBP}}$, this layer could be written as

$$\boldsymbol{\mu}_{\text{FBP}} = (\mathbf{H}_{\text{w}}^{\text{fan}})^{\text{T}} \mathbf{F}\mathbf{W}\hat{\mathbf{p}} \tag{3}$$

Here, $\mathbf{W}$ is a diagonal matrix for weighting, $\mathbf{F}$ represents a ramp filter in detector axis for all views and $\mathbf{H}_{\text{w}}^{\text{fan}}$ is a weighted back-projection operator corresponding to $\mathbf{H}^{\text{fan}}$. All these matrixes can be pre-calculated. In this way, the loss can backpropagate from image domain to projection domain.

**Image Domain CNN.** Finally, we use image domain CNN to suppress noise and improve image quality further. $\boldsymbol{\mu}_{\text{FBP}}$ is the input of image domain CNN and the output is the final reconstruction $\hat{\boldsymbol{\mu}}$.

Overall, the complete cascaded network architecture is shown in Fig. 2. The ground truth of the whole network is corresponding high quality image $\boldsymbol{\mu}^*$ and the loss function for supervised learning is root mean-squared error (RMSE):

$$\text{RMSE} = \sqrt{\frac{1}{m} \sum_{i \in \text{Train dataset}}^{m} (\hat{\boldsymbol{\mu}}_i - \boldsymbol{\mu}_i^*)^2} \tag{4}$$

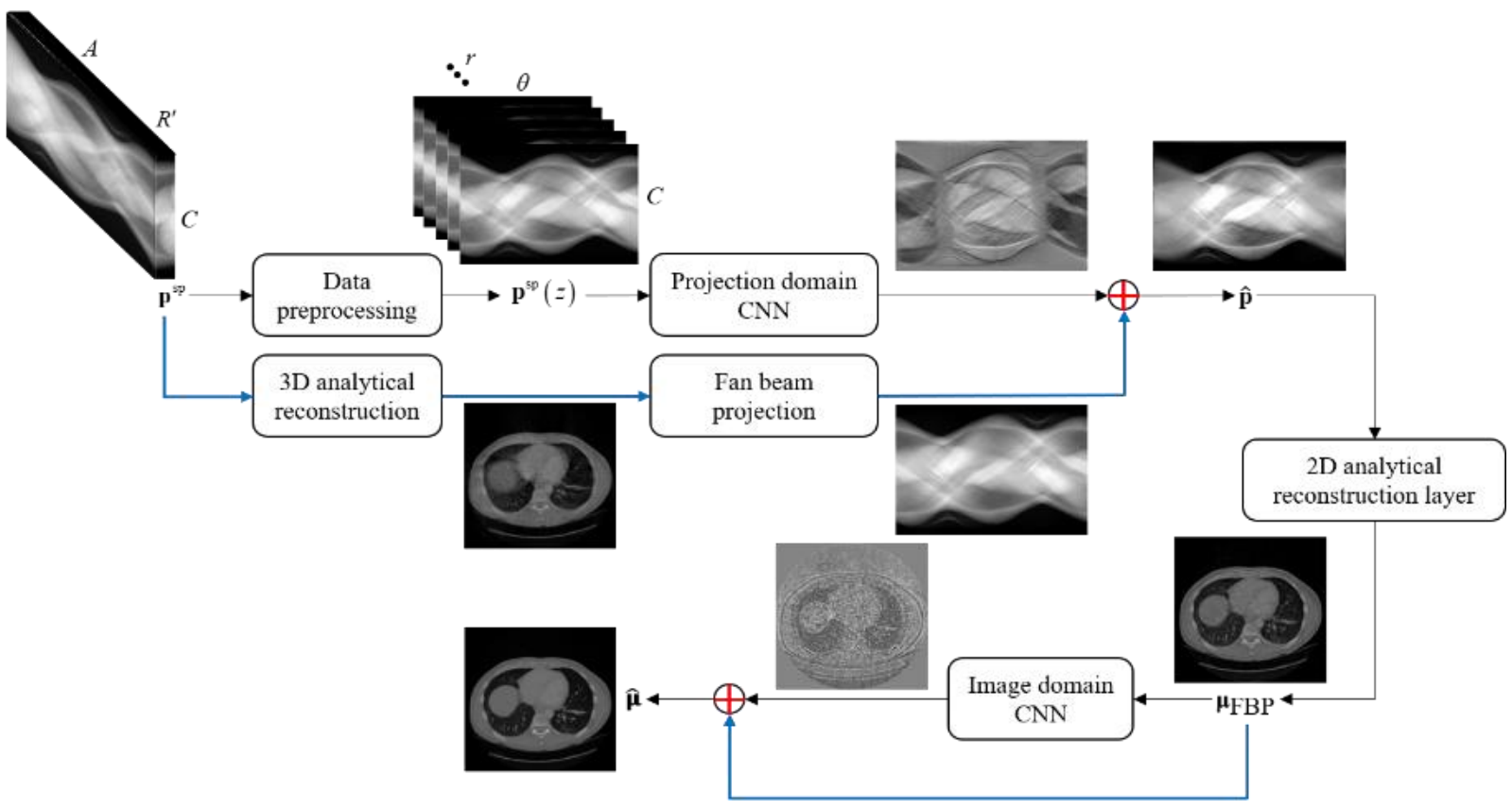

**Fig. 2.** The cascaded network architecture.

## 2.3 Detailed network architecture and training

We use CNN similar to U-Net as backbone in both projection domain and image domain. Fig. 3 shows the detailed parameters. The network in projection domain is adjusted a little deeper than image domain, because the points in projection domain are global related, which means they need a larger receptive field. Both networks are trained to learn the residual between reference and ground truth. In projection domain, we use analytical reconstruction methods, PI-ORIGINAL [6], to reconstruct each slice from helical projection data, and then do fan-beam projection to each slice as the reference. In image domain, we use $\boldsymbol{\mu}_{\mathrm{FBP}}$ as the reference.

We train projection domain CNN first, then we train the whole deep learning reconstruction network end to end.

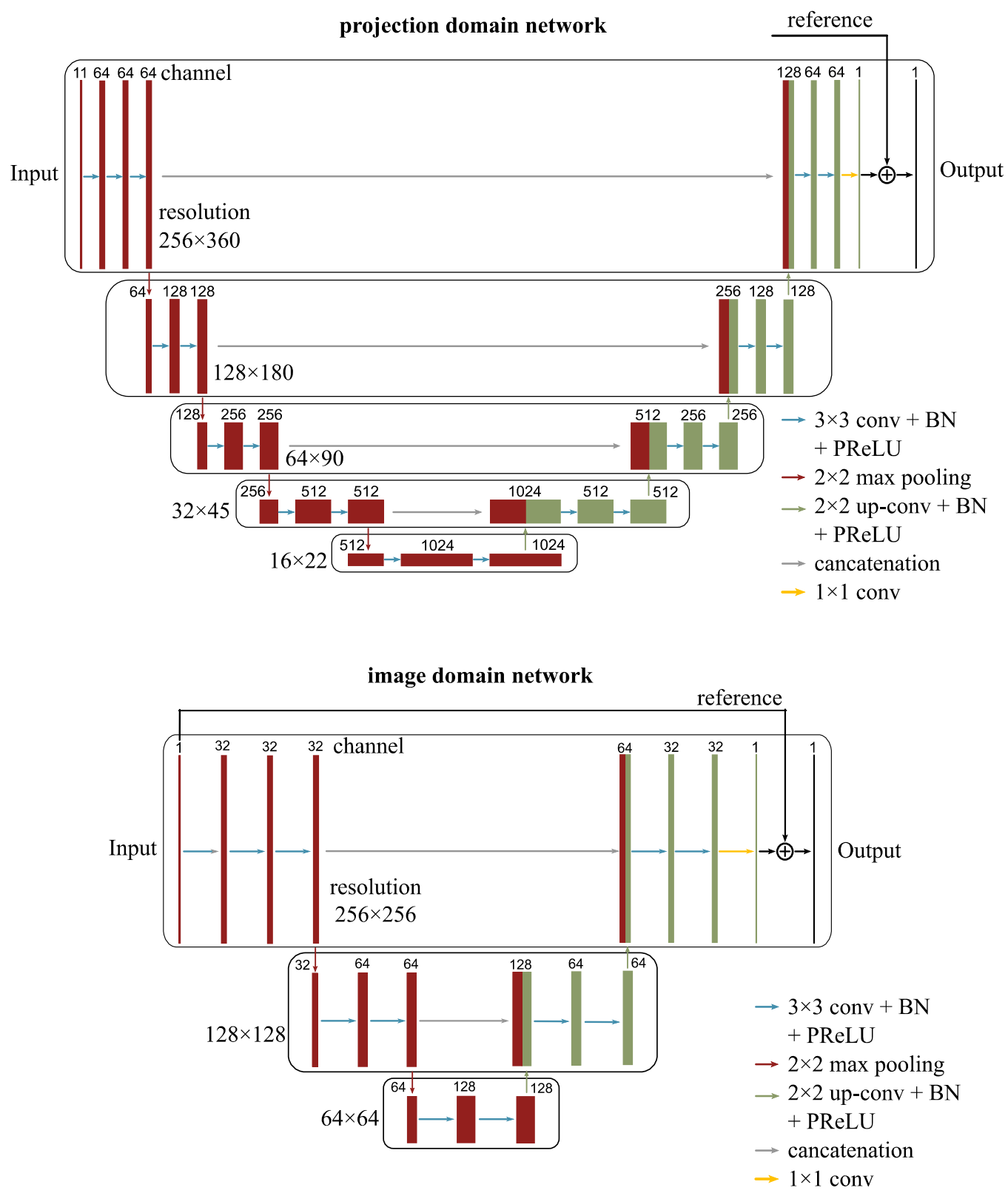

**Fig. 3.** Detailed parameters of CNN

## 3 Experiments and Results

In our simulation experiments, we use lung CT datasets at Give A Scan® [19]. The train, validation and test datasets contain 25, 3 and 3 patients respectively. Each patient has 96 slices and we set different initial scanning angles to do data augmentation. In addition, we add 30000 photons Poisson noise. The detailed geometry parameters are listed in Table 1. In our sparsely spaced multi-detector HCT, the pitch is 1, which means the distance of the object moving in one rotation is equal to the height of detector. And the maximum cone angle is 5.43°.

In training, we use RMSE as loss function and Adam with Nesterov momentum (Nadam) [20] as optimizer with learning rate being 0.0001. We use relative root mean-squared error (RRMSE) and structural similarity index (SSIM) between the output and ground truth for the test dataset to evaluate our method.

$$\text{RRMSE}=\sum_{k\in\text{Test dataset}}\frac{\left\|\hat{\boldsymbol{\mu}}_k-\boldsymbol{\mu}_k^*\right\|_2}{\left\|\boldsymbol{\mu}_k^*\right\|_2} \tag{5}$$

$$\text{SSIM}=\frac{\left(2\bar{\hat{\boldsymbol{\mu}}}\overline{\boldsymbol{\mu}^*}+C_1\right)\left(2\sigma_{\hat{\boldsymbol{\mu}}\boldsymbol{\mu}^*}+C_2\right)}{\left(\bar{\hat{\boldsymbol{\mu}}}^2+\overline{\boldsymbol{\mu}^*}^2+C_1\right)\left(\sigma_{\hat{\boldsymbol{\mu}}}^2+\sigma_{\boldsymbol{\mu}^*}^2+C_2\right)} \tag{6}$$

where $\hat{\boldsymbol{\mu}}$ and $\boldsymbol{\mu}^*$ are the output of the network and the ground truth. $\bar{\hat{\boldsymbol{\mu}}}$ and $\overline{\boldsymbol{\mu}^*}$ are means of $\hat{\boldsymbol{\mu}}$ and $\boldsymbol{\mu}^*$ respectively, $\sigma_{\hat{\boldsymbol{\mu}}}$ and $\sigma_{\boldsymbol{\mu}^*}$ are standard deviations of $\hat{\boldsymbol{\mu}}$ and $\boldsymbol{\mu}^*$, and $\sigma_{\hat{\boldsymbol{\mu}}\boldsymbol{\mu}^*}$ is covariance. $C_1$ and $C_2$ are constants.

**Table 1.** Geometry parameters.

| SO | SD | Detector column | Detec-tor row | Scanning Angles in 360° | pitch | Sparse times | Image size |
|---|---|---|---|---|---|---|---|
| 230 | 400 | 256×0.5 | 16×1 | 360 | 1 | 5 | $(256\times0.25)^2$ |

### 3.1 Results of proposed method on test datasets

We compare our method with PI-ORIGINAL, weighted least square with TV constraint (WLS-TV), and CNN in image domain only. The results are shown in Fig. 4 and Table 2. With our trained network, reconstructing one slice can completed in less than one second. WLS-TV takes much longer time and can't be applied in practice, so we just compare it for reference in this section.

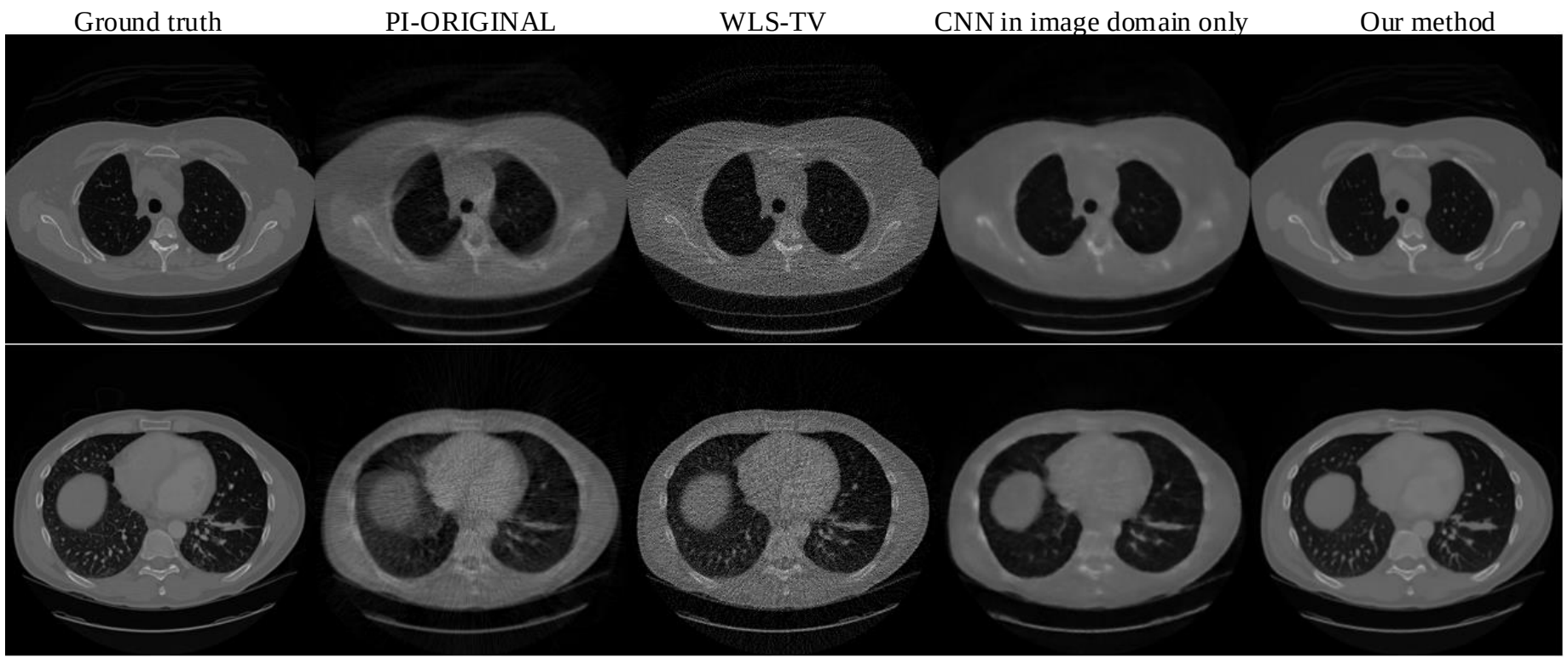

**Fig. 4.** Results on test dataset. Display window: [0, 0.4].

### 3.2 Generalization to different noise level datasets

To examine the generalization capability of our method, we first test the network on simulated lung CT datasets with different noise level. More specifically, our network trained on the noisy dataset of 30000 incident photons per ray, then directly tested on datasets with 20000 and 10000 incident photons per ray. The results are shown in Fig. 5 and Table 3.

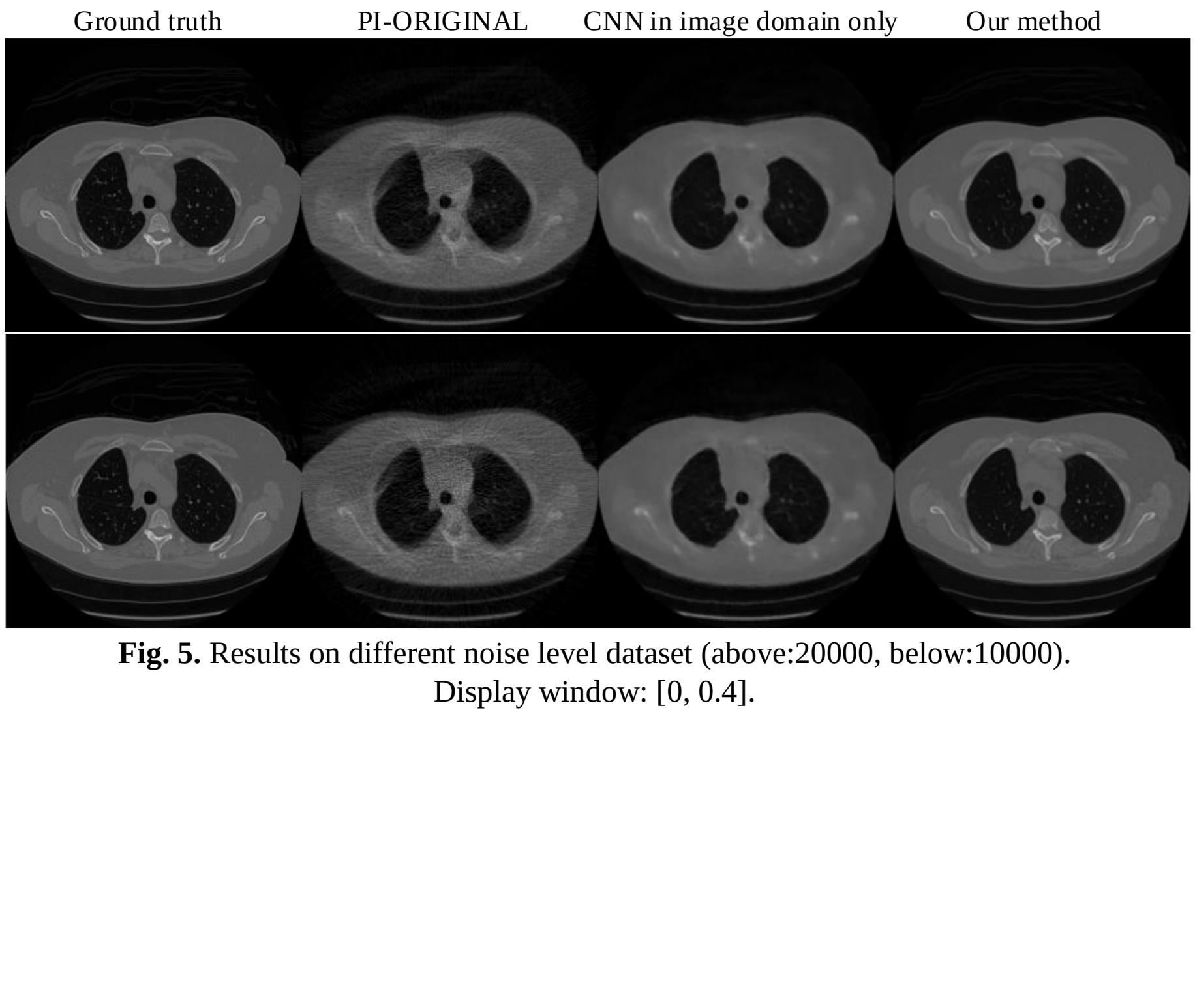

**Fig. 5.** Results on different noise level dataset (above:20000, below:10000). Display window: [0, 0.4].

**Table 2.** Quantitative comparison in experiment 3.1.

| Different methods | RRMSE | SSIM |
|---|---|---|
| PI-ORIGINAL | 22.51% | 94.99% |
| WLS-TV | 7.66% | 99.46% |
| CNN in image domain only | 12.99% | 98.40% |
| Our method | **6.56%** | **99.60%** |

**Table 3.** Quantitative comparison in experiment 3.2.

| Different methods | RRMSE | SSIM |
|---|---|---|
| PI-ORIGINAL | 23.56% (+1.05%) | 94.56% (-0.43%) |
| CNN in image domain only | 13.05% (+0.06%) | 98.36% (-0.04%) |
| Our method | **7.61% (+1.05%)** | **99.47% (-0.13%)** |

### 3.3 Generalization to vastly different datasets

We use dental CT datasets as test dataset in order to further demonstrate the generalization of our method. The dental CT datasets come from our collaborators. The network is also trained with the lung CT dataset only and without any finetuning before test. The results are shown in Fig. 6 and Table 4.

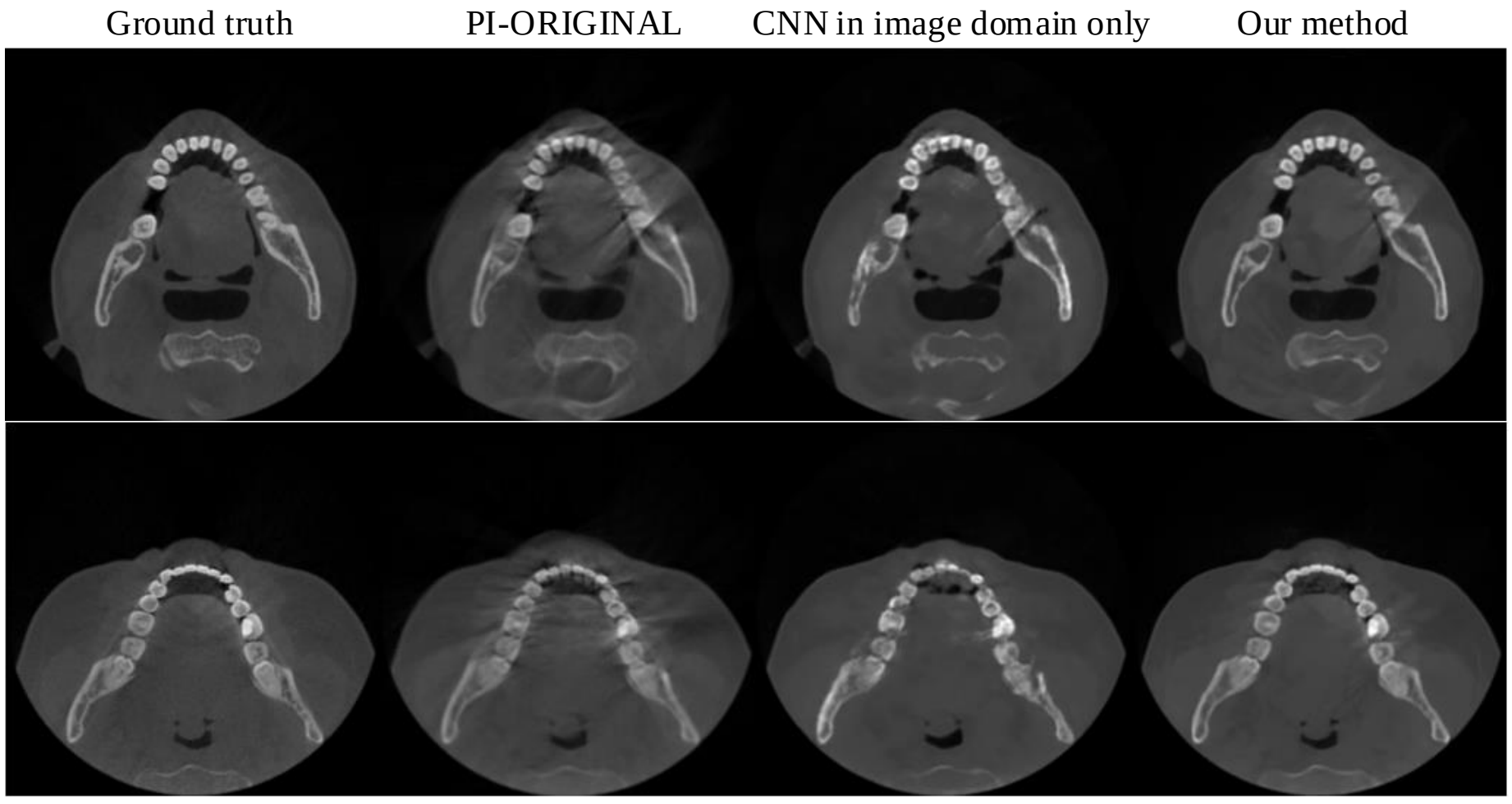

**Fig. 6.** Results on dental CT dataset. Display window: [0, 0.6].

**Table 4.** Quantitative comparison in experiment 3.2.

| Different methods | RRMSE | SSIM |
|---|---|---|
| PI-ORIGINAL | 20.79% (-1.72%) | 95.89% (+0.90%) |
| CNN in image domain only | 19.72% (+6.73%) | 96.53% (-1.87%) |
| Our method | **9.24% (+2.68%)** | **99.26% (+0.34%)** |

## 4 Discussions

From section 3.1, we can conclude that two CNN in dual-domain trained end to end outperforms image domain CNN. Moreover, in our method, we can use image domain's information to adjust the parameters of projection domain CNN. Compared with PI-ORIGINAL and WLS-TV, our method can get a satisfactory result with an affordable computational cost.

Section 3.2 and Section 3.3 shows our method is quite robust to different noise level and vastly different datasets without further training. Image domain CNN is more specific, which means if the test dataset is quite different from train dataset, the image domain CNN will not work perfectly. However, projection data is radon transform of images, and it contains information of CT scanning geometry. Thus, projection domain CNN has a better performance of generalization.

## 5 Conclusions and Future Work

In this paper, we propose a deep learning reconstruction method for sparsely spaced multidetector helical CT which can reduce radiation dose to some extent and cut down the cost of detectors. Our method is based on cascaded dual-domain CNN and the results outperform the conventional reconstruction method in image quality and computational efficiency. From our experience, an effective way to deal with a 3D reconstruction problem is to convert it to 2D. We also show that our method is robust to different noise level. Moreover, our network trained on lung CT datasets can achieve an encouraging result on dental CT datasets without further training, which shows its good potential in generalization. However, our method also has some drawbacks. Compared with lung CT datasets, the RRMSE of dental CT datasets is 2.68% higher, which may be serious in clinical diagnosis. This means we have to collect abundant datasets for training to get a better performance of our method.

The proposed method makes a successful attempt on 3D helical reconstruction using insufficient projection data. In future, we will collect some clinical projection data from helical CT device to further evaluate our method. Also, we will research more about transfer learning for our method.